\documentclass[aps,prb,twocolumn,showpacs]{revtex4}
\usepackage{mathptmx}
\usepackage{bm}
\usepackage{graphicx}

\newcommand{\smeq}{\! = \!}

\newcommand{\smpl}{\! + \!}
\newcommand{\smmi}{\! - \!}

\newcommand{\be}{\begin{equation}}
\newcommand{\ee}{\end{equation}}
\newcommand{\bea}{\begin{eqnarray}}
\newcommand{\eea}{\end{eqnarray}}

\begin{document}

\title{Detection of spin polarized currents in quantum point contacts via transverse electron focusing}
\author{A. Reynoso}
\affiliation{Instituto Balseiro and Centro At\'{o}mico Bariloche, Comisi\'{o}n
Nacional de Energ\'{\i}a At\'{o}mica, (8400) San Carlos de Bariloche, Argentina.}
\author{Gonzalo Usaj}
\affiliation{Instituto Balseiro and Centro At\'{o}mico Bariloche, Comisi\'{o}n
Nacional de Energ\'{\i}a At\'{o}mica, (8400) San Carlos de Bariloche, Argentina.}
\author{C. A. Balseiro}
\affiliation{Instituto Balseiro and Centro
At\'{o}mico Bariloche, Comisi\'{o}n Nacional de Energ\'{\i}a At\'{o}mica, (8400) San
Carlos de Bariloche, Argentina.}
\date{submitted 10 october 2006}

\begin{abstract}
It has been predicted recently that an electron beam can be polarized when it flows adiabatically through a quantum point contact
in a system with spin-orbit interaction. Here, we show that a simple transverse electron focusing setup can be used to
detect such polarized current. It uses the amplitude's asymmetry of the spin-split
transverse electron focusing peak to extract information about the electron's spin
polarization. On the other hand, and depending on the quantum point contact geometry, including this one-body effect can be important when using the focusing setup to study many-body effects in quantum point contacts.
\end{abstract}
\maketitle

\section{Introduction}
Important goals in the field of spintronics\cite{Spintronicsbook} are the
production, detection and manipulation of spin polarized currents in
semiconductors. During the last years, there has been several proposals in
those directions which go from the use of ferromagnetic materials\cite
{DattaD90} to the use of the spin Hall effect in systems with spin-orbit (SO)
interaction.\cite{DyakonovP71,Hirsch99,EguesBL03,MurakamiNZ03,SinovaCSJM04,MolenkampSB01,SihMKLGA05,EngelHR05,SihLMHGA06} The
latter approach has attracted much of attention due to the non-trivial spin
dynamics introduced by the SO coupling.\cite{MishchenkoH03,SchliemannL03,SchliemannLW05,NikolicZW05} In a recent work by Eto \textit{et al%
}\cite{EtoHK05} a very simple mechanism for creating a spin polarized
current was proposed: electrons in a two dimensional gas (2DEG) with SO
coupling can be polarized when they pass adiabatically through a constriction that admits
only a few conducting channels. Such a constriction could be a quantum point
contact (QPC) or simply a potential barrier in a quantum wire---a similar
effect was predicted by Silvestrov and Mishchenko.\cite{SilvestrovM05} The origin of
such polarization, discussed below, is related to the spin structure of the energy sub-bands
inside the constriction or barrier.
Despite this seemingly simple way to create a spin polarized current, this effect has not been observed yet. One of the reasons is that its detection requires the use of a spin analyzer, which is not a trivial task.

In this work, we show that a two terminal device is not efficient to detect such spin
polarized current and propose to use transverse electron focusing to measure it.
Transverse focusing experiments are done in a
solid state device in which electrons emitted from a QPC are focalized
onto a collector (another QPC) by the action of an external magnetic field
perpendicular to the 2DEG---in a classical picture, the electrons are forced to follow circular orbits due to the Lorentz force. For values of the external field $B_n$ such that the distance between the QPCs is $n$ times the diameter of the cyclotron orbit, with $n$ an integer number, the electrons enter the collector and create a charge accumulation in it that generates a voltage difference. This gives voltage peaks as the external field is swept through the focusing fields $B_n$.\cite{PotokFMU02} As shown in Refs. [%
\onlinecite{UsajB04_focusing,RokhinsonLGPW04}], in systems with spin-orbit
coupling the first focusing peak splits in two. Furthermore, each peak
corresponds to a different spin projection of the electron leaving the
emitter.\cite{UsajB04_focusing} If the emitted electrons were unpolarized,
both peaks would have the same intensity. Conversely, if the electrons
leaving the emitter are polarized, the intensity of the two peaks must be
different. The intensity difference of the two peaks is then a direct
measurement of the spin polarization of the current induced by the emitter.
This very simple idea was recently used in Ref. [\onlinecite{RokhinsonPW06}]
to study the current's spin polarization induced by the `$0.7$' anomaly in a
QPC. Our results are also relevant in that context, since they shown that
one-body effects can substantially contribute to the peak's height asymmetry. It might then be difficult to distinguish one-body from many-body effects\cite{RejecM06,CornagliaB04,KindermannBM06} using this technique.

\section{Spin polarization in a quantum point contact}
\subsection{The model}
\begin{figure}[t]
\includegraphics[width=.44\textwidth,clip]{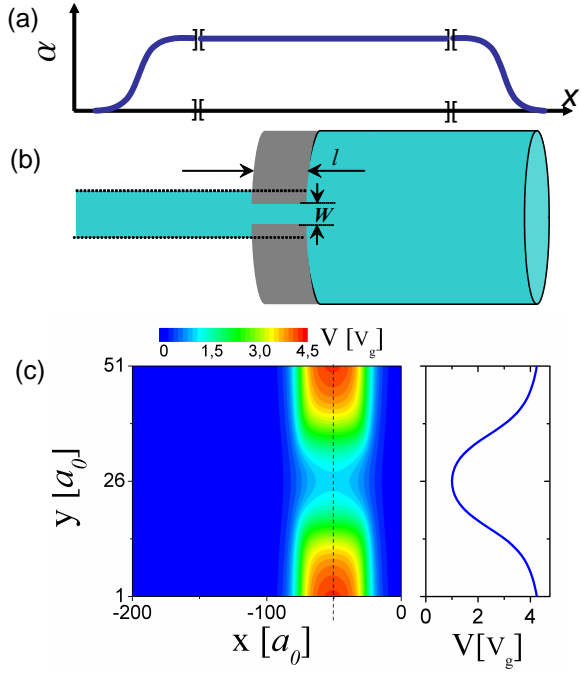}
\caption{(color online) (a) Schematic representation of the spatial profile of the SO coupling $\alpha$. The coupling is turned on and off in $300$ lattice sites. (b) Schematics of the system's geometry. The period lead (800 sites wide) at the right of the QPC is introduced to avoid the oscillation of the polarization as a function of the sample-lead interface (see text). (c) potential profile corresponding to Eq. (\ref{Ando})  }
\label{fig0}
\end{figure}

We consider a 2DEG with Rashba spin-orbit coupling. The Hamiltonian
describing the system is given by 
\begin{equation}
H=\frac{p_{x}^{2}+p_{y}^{2}}{2m^{*}}+\frac{\alpha }{\hbar }(p_{y}\sigma
_{x}\!-\!p_{x}\sigma _{y})+V(\mathbf{r})\,.  \label{ham}
\end{equation}
The first term is the kinetic energy of the 2DEG where $m^{*}$ is the electron
effective mass. The second term describes the Rashba coupling, whose
strength is characterized by $\alpha $, and the last term is the lateral
confining potential defined by the external gates. In order to present a
quantitative description for arbitrary geometries, the Hamiltonian is
integrated numerically using a finite difference scheme---this is equivalent
to work with a tight-binding model (see Ref. [\onlinecite{ReynosoUB06}] for
details). We first analyze the electronic transport through a barrier or a
constriction. The confining
potential is $V({\bm{r}}_{n})\!=\!V_{s}({\bm{r}}_{n})\smpl V_{c}({\bm{r}}_{n})$, where $V_{s}({\bm{r}}_{n})$ defines the shape of the sample (a hard wall potential) and $V_{c}({\bm{r}}_{n})$ the structure of the barrier or QPC. In
what follows, we use two different models for $V_{c}$ and analyze their
effectiveness to spin-polarize the transport current. The SO coupling $%
\alpha $ is set to zero at the source and drain reservoirs---described as ideal
leads---and it is turned on adiabatically at the lead-sample interface using either a hyperbolic  tangent function or a simple linear function to describe its spatial profile (see Fig. \ref{fig0}a).\cite{ReynosoUB06,ReynosoUB06_Lawn,note1}

In the linear response regime, the conductance is given by $G\! = \!(e^{2}/h)%
\mathrm{Tr}\left[\bm{\Gamma }^{R}\bm{G}^{r} (E_{\text{F}})\bm{\Gamma }^{L}%
\bm{G}^{a} (E_{\text{F}})\right]\,, $ where $\bm{G}^{r(a)}(\varepsilon)$ is
the retarded (advanced) matrix propagator, with elements $\bm{G}_{i\sigma
,j\sigma ^{\prime }}^{r}(\varepsilon)$ given by the propagator from site $j$
and spin $\sigma^{\prime}$ to site $i$ and spin $\sigma$, $\bm{\Gamma }%
_{i\sigma ,j\sigma ^{\prime }}^{L(R)}\! = \!{\mathrm{i}}(\bm{\Sigma}%
_{L(R)}^{r}\! - \!\bm{\Sigma}_{L(R)}^{a})_{i\sigma ,j\sigma ^{\prime }}$
where $\bm{\Sigma}_{L(R)}^{r(a)}$ is the retarded (advanced) self-energy due
to the left (right) contact and $E_{\text{F}}$ is the Fermi energy. The SO
coupling acts as an effective magnetic field contained in the plane of the
2DEG with a magnitude that is proportional the momentum of the carrier. This
effective field lifts the spin degeneracy of the bands. For the wire
geometry it is convenient to quantize the spins along the transverse axis ($%
y $-axis). In what follows $\uparrow$ and $\downarrow$ indicate the two spin
projections along this direction.

To study the current spin polarization, we calculate the spin-resolved
conductance $G_{\sigma ^{\prime }\sigma }$ that describes the contributions
due to the electrons that are injected from the left lead with spin $\sigma $
and collected at the right lead with spin $\sigma ^{\prime }$. The
polarization $P$ of the transmitted current is then defined as $
P\!=\!\sum_{\sigma }(G_{\uparrow \sigma }\!-\!G_{\downarrow \sigma
})/\sum_{\sigma \sigma ^{\prime }}G_{\sigma ^{\prime }\sigma }$. We first
consider the following potential\cite{EtoHK05} 
\bea
\label{Ando}
V_{c}(x,y)\!&=&\!\frac{V_{g}}{2}(1\!+\!\cos \frac{\pi x}{L_{x}})\\
\nonumber
&&+\frac{%
\eta E_{\text{F}}}{\Delta ^{2}}\sum_{s=\pm }(y\!-\!y_{s}(x))^{2}\theta
[s(y\!-\!y_{s}(x))]\,,
\eea
which is defined for $|x|\!\leq \!L_{x}\!=\!L_{1}\,\theta
(-x)\!+\!L_{2}\,\theta (x)$ and it is zero otherwise. $\!L_{1}+\!L_{2}$ is
the total length of the potential barrier, $y_{\pm }(x)\!=\!\pm
(L_{y}/4)[1\!-\!\cos (\pi x/L_{x})]$ gives the shape of the lateral constriction, and $\theta (x)$
is the step-function. This model potential has the virtue that the relevant
parameters can be easily changed. The gate potential $V_{g}$ controls the
height of the potential barrier at $x\!=\!0$ and then the conductance. 
\begin{figure}[t]
\includegraphics[width=.48\textwidth,clip]{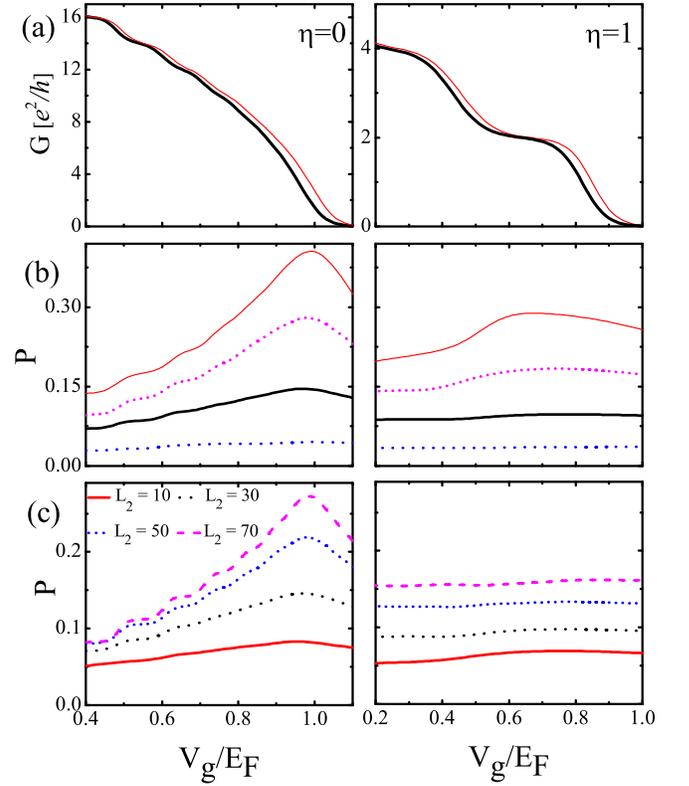}
\caption{(color online) Conductance (a) and polarization (b, c) for
different geometries ($\eta \!=\!0$, left and $\eta \!=\!1$, right  panel) and
values of the SO-coupling $\alpha $ as a function of the gate potential $%
V_{g}$. The common parameters are $L_{y}\!=\!51a_0$, $L_{1}\!=\!L_2\smeq30a_{0}$, $%
E_{\text{F}}\!=\!10$meV, $m^{*}\!=\!0.067m_{0}$, $\Delta \!=\!L_1/4$ and the
lattice parameter $a_{0}\!=\!5$nm. (a) conductance for $\alpha \!=\!5$meVnm
(thick line) and $20$meVnm (thin line); (b) polarization for $\alpha
\!=\!5 $ (dotted line), $10$ (solid line), $15$ (shot-dotted line), and $20$meVnm
(thin solid line); (c) polarization for $\alpha \!=\!10$meVnm and
different values of $L_{2}$.}
\label{fig1}
\end{figure}

Before presenting the results, it is important to emphasize here the role play by the sample-lead interface (the point where $\alpha$ is turned off). We have found that, in narrow wires, the polarization of the current strongly depends on the position of such interface. In fact, the polarization shows an oscillating pattern as a function of the distance between the QPC and the sample-lead interface. The oscillation originates from the scattering of the electrons at the edges of the wires before reaching the sample-lead interface. To  avoid this effect, we introduced a large periodic lead at the output of the QPC as shown in Fig. \ref{fig0}b.\cite{note2} 

Figure \ref{fig1} shows the conductance and the polarization as a function
of the gate potential $V_{g}$ for different values of the parameters. Left
panels correspond to $\eta \!=\!0$, a simple potential barrier, while the
right panels correspond to $\eta \!=\!1$, a QPC. In Fig.\ref{fig1}a the
conductance is shown for different values of $\alpha $; as it increases the
conductance curve is shifted towards a higher gate potential, as expected. Except for that, the total
conductance is not sensitive to the Rashba coupling. Note that for $\eta\smeq0$ the conductance steps are not well defined. It is worth notice that the latter case is not in the regime where a jump on the conductance at the opening of the barrier is expected.\cite{SilvestrovM05}
Figure \ref{fig1}b
shows the current polarization for $\alpha \!=\!5,10,15,20$ meVnm. The
 polarization increases with $\alpha $ monotonically.
The polarization for $\alpha \!=\!10$ meVnm and different values of $L_{2}$
is shown in Fig. \ref{fig1}c. Clearly, $P$ increases with $L_{2}$ while it
is essentially independent of $L_{1}$ (not shown).\cite{EtoHK05}

These results are consistent with the ideas put forward in Ref. [%
\onlinecite{EtoHK05}]. The spin filtering effect arises from the avoided
crossings (caused by the term $\frac{\alpha}{\hbar×} p_y \sigma_x$ in (1)) between different spin-dependent sub-bands (channels associated to the term $p_x^2/2m^*\smmi \frac{\alpha}{\hbar×} p_x \sigma_y$). These
avoided crossings generate an adiabatic spin rotation as the electrons leave
the QPC, hence the strong dependence on $L_{2}$ (the parameter that controls
adiabaticity). The picture also explains why the total conductance is not
affected, as the number of propagating channels inside the QPC do not change.
The partial conductance $G_{\uparrow }=\sum_{\sigma }G_{\uparrow,\sigma
}$ and $G_{\downarrow }=\sum_{\sigma }G_{\downarrow ,\sigma}$ are, in general,
very different from each other. Usually, in the first plateau, $%
G_{\uparrow }$ is larger than $e^{2}/h$, which indicates that for the
transmitted electrons with spin $\uparrow $ there is more than one channel
that contributes. In particular, one can find cases where $P\!\simeq \!1$, so
that $G_{\downarrow \downarrow }\simeq G_{\uparrow \downarrow }\simeq
G_{\downarrow \uparrow }\simeq 0$ and $G_{\uparrow \uparrow }\simeq
2e^{2}/h$ (see Ref. [\onlinecite{ReynosoUB06_Lawn}]). In this extreme case,
only spin up electrons are transmitted. However, spins rotate as
electrons enter and leave the QPC and both spins contribute to the charge
current inside the QPC.

\begin{figure}[t]
\centering
\includegraphics[height=.67\textwidth,clip]{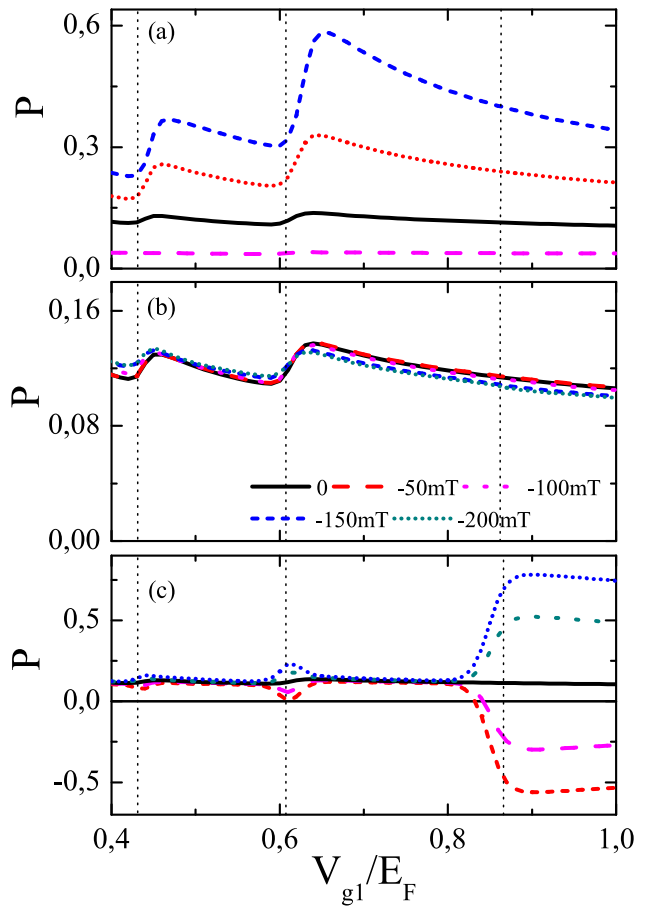}
\caption{(color online) Polarization as a function of $V_g$ in a more
realistic potential profile (see text). The vertical lines correspond to $%
G\! = \! e^2/h$ and $G\! = \! 3 e^2/h$. (a) for $\alpha\! = \!5,10,15,20$%
meVnm (dashed, solid, dotted, short-dashed line, respectively). (b) for
different values of $B_z$ and $\alpha\! = \!10$meVnm. (c) for different
values of the Zeeman energy $g\mu_B B_\parallel\smeq-0.25$meV, $-0.125$meV, $0$meV, $0.125$meV, $0.25$meV (short-dashed, dashed, solid, dotted and short dotted line, respectively) and $\alpha$ as in (b). Notice that for $%
B_\parallel\! \neq \!0$, the polarization has an abrupt change when the
conductance reaches the first plateau and that for $B_\parallel\!<\!0$ it
can even change sign. }
\label{fig2}
\end{figure}

The fact that $P\!\neq \!0$ is quite general. It does not depend on the
details of the potential $V_{c}(\bm{r})$ as far as adiabaticity is
guarantee. We have verified this by computing the conductance and the
polarization for different potential profiles. From hereon, we use a more
realistic potential corresponding to rectangular gates of length $l$, separated a distance $W$ from each other and 
located at a distance $z$ from the 2DEG, \cite{Ferrybook_potential}
\bea
\nonumber
V(x,y) &\smeq& V_{g}[f(x_{-},y_{+})\smmi f(x_{+},y_{+})\\
&&+f(x_{-},-y_{-})-f(x_{+},-y_{-})]
\eea
with $x_{\pm }=x/z\pm l/2z$, $y_{\pm }=y/z\pm W/2z$ and 
\bea
\nonumber
f(u,v)&=&\frac{1}{2\pi }\left[\frac{\pi }{2}-\arctan (u)-\arctan (u)\right.\\
&&\left.+\arctan \left(\frac{uv}{\sqrt{1+u^{2}+v^{2}}}\right)\right]
\eea
In the following, we use  $z\!=\!30$nm, $l\smeq250$nm and $W\smeq100$nm

\subsection{The effect of magnetic fields}
We now analyze the effect of both a small
out-of-plane magnetic field, which couples to the spin and orbital degrees of
freedom, and an in-plane field, which couples to the spins only. As we show below,
the polarization $P$ is more sensitive to the presence of an in-plane field.

The presence of an out-of-plane field $B_{z}$ is described by including  a term
$H_{z}\! = \! g\mu_{B}\sigma _{z}B_{z}$ and replacing the
momentum $\bm{p}$ by $(\bm{p}\! + \! e/c\bm{A})$ where $\bm{A}$ is the
vector potential associated with $B_{z}$. In the tight-binding
representation, the latter is included as a Peierls
substitution.\cite{UsajB04_focusing} Both the conductance (not shown) and the polarization are
relatively insensitive to the presence of $B_z$ in the explored range $%
[0,200 $mT$]$. In the case of the conductance, the field does not generate a splitting of the plateaus but a weak double shoulder. This effect, which is much larger than what it is expected from the Zeeman term, is originated by the diamagnetic coupling.\cite{DebaldK05}

\begin{figure}[t]
\centering
\includegraphics[height=.25\textwidth,clip]{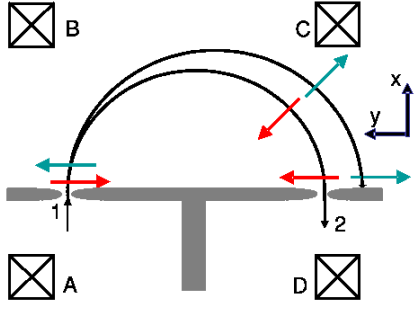}
\caption{(color online) Schematic representation of the setup used to detect
spin polarization. The focusing signal is the voltage $V_{CD}$ as a function
of $B_z$}
\label{fig3}
\end{figure}

The case of an in-plane field along the $y$-axis is shown in figure \ref
{fig2}c. Since the transmitted current is polarized in the $y$-direction, a
field pointing in the same direction tends to increase or decrease the
effect depending on its sign. This is clearly seen in the figure, in
particular before the first plateau of the conductance is fully developed---the effect is less pronounced at the opening of the second and third channels.
Interestingly, for the parameters of the figure and a field such that $g\mu
_{B}B_{\parallel}\!=\!0.125$meV, which for $g\smeq0.5$ corresponds to $B_\parallel\!\simeq\!5$T, $P$ changes sign for a gate potential $V_{g}$ corresponding
to $G\!\simeq \!e^{2}/h$. This shows that the interplay between SO coupling
and an external in-plane field could be used to select the desired
polarization of the current transmitted through a QPC.

\section{Detection using transverse electron focusing}
Even in the presence of external fields, a direct measurement of transport
properties in two terminal devices does not provide any evidence of the spin
polarization of the transmitted current. We have verified that neither the shot
noise nor the thermopower show any significant feature even in the
case of large $P$. To measure $P$ it is then necessary to design an
experiment with a more complex geometry. In the following, we discuss how
transverse electron focusing can give a direct measurement of the
polarization induced by the QPC.

The transverse electron focusing setup is shown in Fig. \ref{fig3}:  a
current $I_{AB}$ is injected  through the first QPC while the voltage drop
$V_{CD}$ is measured in the second QPC.\cite{vanHouten89} $V_{CD}$ shows then a series
of peaks every time $R/2r_{c}$ is an integer number. Here, $R$ is the distance
between the QPCs (of the order of $1\mu $m) and $r_{c}$ is the cyclotron
radius. For the purpose of the present work we can assume\cite{BeenakkerH91}
that $V_{CD}\!\propto \!T_{AD}$, where $T_{AD}$ is the transmission probability
from one QPC to the other. In the presence of SO coupling, the first
focusing peak is split in two, each peak corresponding to a different spin
orientation of the electrons leaving the emitting QPC.\cite
{UsajB04_focusing,RokhinsonLGPW04,ReynosoUSB04,Heremans,DedigamaDMGKSSMH06}

\begin{figure}[t]
\centering
\includegraphics[height=.67\textwidth,clip]{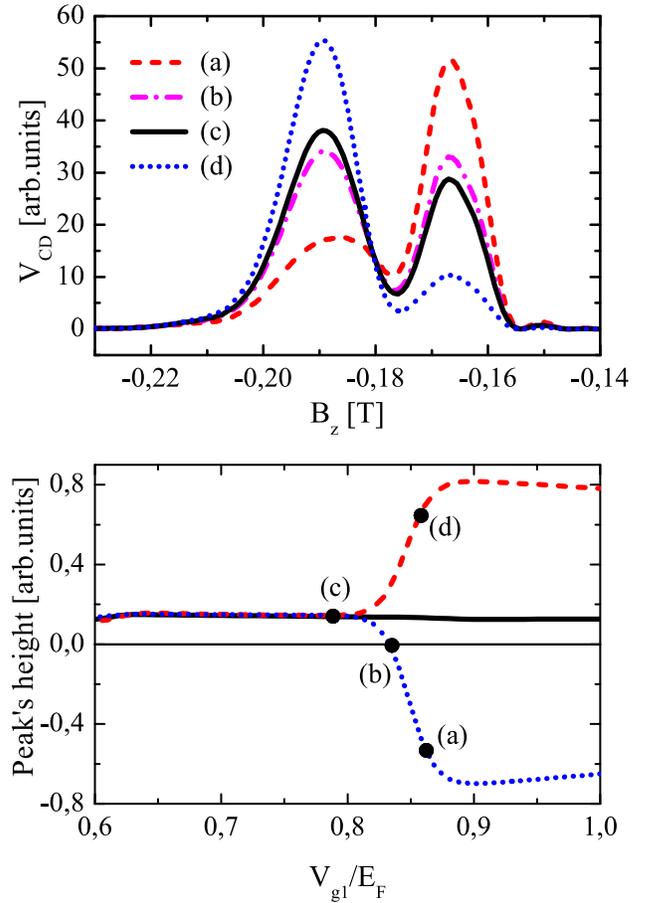}
\caption{(color online) Top: Spin-split first focusing peak for different
values of $V_{g1}$ and $B_\parallel$ as indicated. Peaks are normalized to
have the same area. Notice the contrast in the peaks' magnitude, reflecting
the different values of polarization of the electrons emitted by the QPC.
Bottom: Polarization, as extracted from the peaks' height, as a function of $%
V_{g1}$ for $\alpha\! = \!10$meVnm and $g\mu_B B_\parallel\! = \!0,0.25$meV
and $-0.25$meV (solid, dashed and dotted lines, respectively). These results
compare nicely with those of Fig. \ref{fig2}c and shown that information
about $P$ can be extracted from the focusing signal.}
\label{fig4}
\end{figure}

Figure \ref{fig4}a shows typical results for $T_{AD}$ as a function of $B_z$
for different values of $V_{g1}$ and $B_\parallel$. Clearly, the asymmetry
of the peaks height results from the polarization of the electrons as they
are transmitted through the first QPC. Therefore, such difference is a
measure of $P$. Figure \ref{fig4}b shows the peak's height difference
(normalized to the sum) as a function of $V_{g1}$ for $\alpha\! = \!10$meVnm
and $g\mu_B B_\parallel\! = \!0,0.25$meV and $-0.25$meV. This clearly
reveals the presence of a spin polarized current and allows to quantify it.
A close comparison with Fig. \ref{fig2}c, however, show some quantitative
differences. The main difference comes from the fact that in the focusing
geometry both QPCs lead to some spin polarization. A simple estimate for 
the `measured' polarization $P_m$ is given by $P_m\smeq(P_1\smmi D)/(1\smmi P_1 D)$ where $P_1$ is the spin polarization introduced by
the first QPC and $D\! = \!\sum_{\sigma }(G_{\sigma\uparrow}\! - \!
G_{\sigma\downarrow})/\sum_{\sigma\sigma ^{\prime
}}G_{\sigma^{\prime}\sigma} $ is a filtering factor that measure the transmission difference of the two spin projections through the second QPC (for a symmetric
QPC, $D\! = \!-P$). We checked this relation by turning on and off the SO
coupling in the second QPC. Experimentally, this could be avoided by using a
different geometry for the second QPC (less adiabatic) so that $D\!\ll\!1$
and then $P_m\! \simeq\!P_1$. 

It is important to emphasize that the behavior of $P$ as a function of $%
V_{g1}$ in the region where the QPC starts conducting is, in general, the
opposite to the one observed in a sample with a two dimensional hole gas, see Ref. [\onlinecite{RokhinsonPW06}]. That is,
the polarization increases as the QPC opens up---there are some examples of
a decrease of the polarization (see Fig. \ref{fig1}) though the changes of P is
rather small. This increase of $P$ is also expected based on the theoretical
arguments explained above for either Rashba or Dreselhauss SO coupling.

\section{Summary}

We showed that the spin polarized current generated by a QPC in systems with Rashba spin-orbit coupling can be measured using a transverse electron focusing setup. This is possible because the difference in  amplitude of the spin-split focusing peaks is proportional to the spin polarization of the electrons leaving the emitting QPC. 

In addition, we also showed that the interplay between SO coupling and an in-plane magnetic field could
be used to independently select the desired sign of the polarization of the current transmitted through a given QPC. This is quite different for the case without SO,\cite{PotokFMU02} where only the magnitude of the polarization can be changed and where the sign of the polarization would be the same in all the QPCs present on the system.

Finally, we found that current polarization increases with the conductance of the QPC as it goes from zero to $2e^2/h$. As mentioned above, this is just the opposite behavior that was observed in Ref. [\onlinecite{RokhinsonPW06}]. It is not clear to us at this point what is the origin of this discrepancy. One possibility is the different nature of the spin-orbit coupling in electron and hole gases. The other is the presence of many-body effects, which are not included in our calculation. In the latter case, the temperature dependence of the observed polarization might provide a way to distinguish between both contributions as we expect the one-body SO effect described here to be rather insensitive to temperature as far a $kT$ is smaller than $2\alpha k_F$.
This is a very interesting point that deserves further investigation.

\section{Acknowledgement}
This work was supported by ANPCyT Grants No 13829 and 13476 and CONICET PIP
5254. AR and GU acknowledge support from CONICET.

%\bibliographystyle{apsrev}
%\bibliography{../../bibliography/usaj_1,../../bibliography/usaj_2_jun06,notes}

\begin{thebibliography}{35}
\bibitem[{not({\natexlab{b}})}]{note2}
\bibinfo{note}{In this geometry, there is still a small decaying oscillation
  around the mean value which arises from the the multiple reflexion between
  the sample-lead interface and the hard wall potential that describes the QPC
  far from its center. In this case the oscillation have a well defined
  wavelength, $\lambda\smeq \pi \hbar^2/m^*\alpha$ so it can be easily averaged
  out.}
\end{thebibliography}

\end{document}